\documentclass[a4paper,11pt]{article}
\usepackage{pos}
\newcommand{\gev}{{\rm GeV}}

\title{Neutrinos from charm: forward production at the LHC and in the atmosphere}
 \ShortTitle{Neutrinos from charm: LHC and in the atmosphere}

\author*[a]{Yu Seon Jeong}
\author[b]{Weidong Bai}
\author[c]{Milind Diwan}
\author[d]{Maria Vittoria Garzelli}
\author[c]{Fnu Karan Kumar}
\author[e]{Mary Hall Reno}

\affiliation[a]{Chung-Ang University, High Energy Physics Center\\
Dongjak-gu, Seoul 06974, Republic of Korea}

\affiliation[b]{Sun Yat-sen University,  School of Physics, \\
 No. 135, Xingang Xi Road, Guangzhou, 510275, P. R. China}

\affiliation[c]{Brookhaven National Laboratory, \\
Upton, New York, USA}

\affiliation[d]{Universit\"at Hamburg, II Institut f\"ur Theoretische Physik,\\
Luruper Chaussee 149, D-22761, Hamburg Germany}

\affiliation[e]{University of Iowa, Department of Physics and Astronomy,\\
Iowa City, Iowa 52242, USA\vskip 0.1in}

\emailAdd{yusjeong@cau.ac.kr}
\emailAdd{baiwd3@mail.sysu.edu.cn}
\emailAdd{diwan@bnl.gov}
\emailAdd{garzelli@mail.desy.de}
\emailAdd{fkarankum@bnl.gov}
\emailAdd{mary-hall-reno@uiowa.edu}

\abstract{Theoretical predictions of the prompt atmospheric neutrino flux have large uncertainties associated with charm hadron production, by far the dominant source of prompt neutrinos in the atmosphere. The flux of cosmic rays, with its steeply falling energy spectrum, weights the forward production of charm in the evaluation of the atmospheric neutrino flux at high energies. The current LHCb experiment at CERN constrains charm production in kinematic regions relevant to the prompt atmospheric neutrino flux. The proposed Forward Physics Facility has additional capabilities to detect neutrino fluxes from forward charm production at the LHC. We discuss the implications of the current and planned experiments on the development of theoretical predictions of the high energy atmospheric neutrino flux.}

\FullConference{37$^{\rm{th}}$ International Cosmic Ray Conference (ICRC 2021)\\
		July 12th -- 23rd, 2021\\
		Online -- Berlin, Germany}

\begin{document}
\maketitle

\section{Introduction}
%
%
%
The Large Hadron Collider (LHC) is a powerful source of neutrinos. 
In $pp$ collisions at the LHC, a large number of neutrinos are produced in the forward direction. 
Two experiments, FASER$\nu$ \cite{Abreu:2019yak} and SND@LHC \cite{Ahdida:2020evc} will be carried out to measure the interactions of these forward neutrinos during the Run 3.  
Possible upgrades of these experiments and further additional experiments are foreseen in a next stage, at the so-called Forward Physics Facility at the LHC.  
The high energy of the LHC abundantly produces
neutrinos from the decay of heavy flavor hadrons (prompt neutrinos), 
even in the very large rapidity region ($y \gtrsim 6.5$).

Prompt neutrinos can also be produced in the atmosphere. 
Various secondary hadrons from the incident cosmic ray interaction with air nuclei produce neutrinos in their decays, known as atmospheric neutrinos, over broad energy range. 
While most atmospheric neutrinos are from $\pi^\pm$ and $K^\pm$ decays (conventional neutrinos) and their spectrum covers relatively low energies, at energies of $E_\nu \gtrsim $ 1 PeV, the prompt component of atmospheric neutrinos dominates and plays a role of the main background to astrophysical neutrinos probed by high energy neutrino observatories such as IceCube and KM3NeT. 

The prompt atmospheric neutrino fluxes have large uncertainties, and one of the main reason is due to the poor understanding of heavy flavor production. 
The collision energy of $\sqrt{s}= 13 - 14 {\ \rm TeV}$ corresponds to $E_p$ of the order of $\mathcal{O}$($10^2$ PeV) in the laboratory frame, which is the relevant energy for astrophysical neutrinos and prompt atmospheric neutrinos. 
Therefore, measurements of heavy flavor and prompt neutrino production at forward LHC experiments can contribute to reduce the uncertainty in the theoretical predictions of the prompt atmospheric neutrino flux.

In this work, we evaluate the charm meson production cross section at next-to-leading order (NLO) in perturbative QCD (pQCD) using the PROSA parton distribution functions (PDFs) \cite{Zenaiev:2019ktw} with parameters that make the results well-matched to the measurements by the LHCb experiment \cite{Aaij:2015bpa}. 
We examine the regions of collision energy $\sqrt{s}$ and collider frame rapidity $y$ relevant to prompt atmospheric neutrino flux predictions. We also consider the impact of the small-$x$ and large-$x$ PDFs.

\section{Charm production and PDF} 

 \begin{figure}
    \centering
       \includegraphics[width=.5\textwidth]{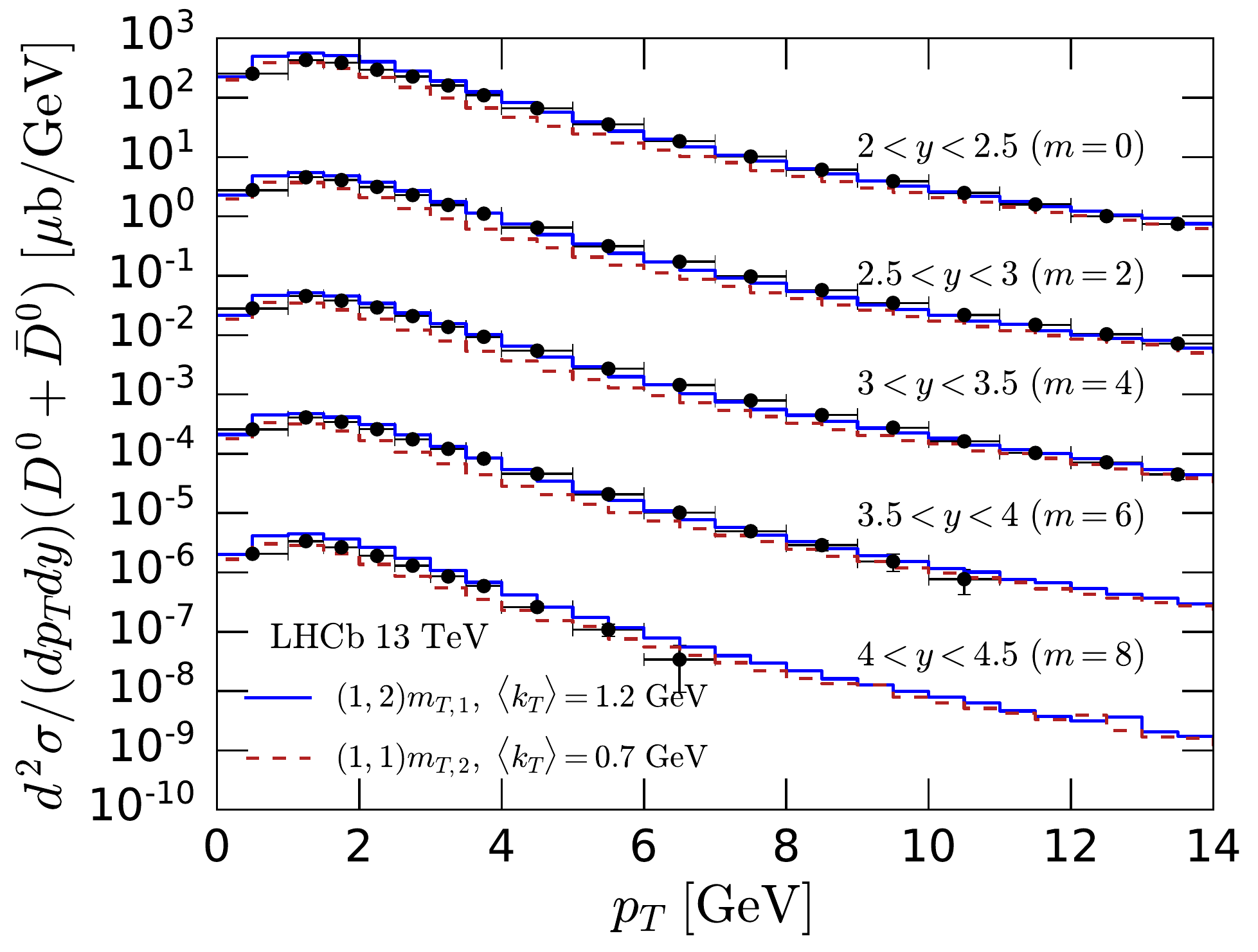}
       \caption{
      The double-differential distribution $d^2\sigma/dp_T\, dy$ for  $D^0+\bar{D}^0$ production at $\sqrt{s}=13$ TeV with (a) renormalization and factorization scales
      $(\mu_R,\mu_F)=(1,2)\cdot (p_{T,c}^2+m_c^2)^{1/2}$ in association with intrinsic $\langle k_T\rangle = 1.2$ GeV (solid)
      and with (b) $(\mu_R,\mu_F)=(1,1)\cdot(p_{T,c}^2+4 m_c^2)^{1/2}$ in association with $\langle k_T\rangle = 0.7$ GeV (dashed), 
      compared with LHCb data  \cite{Aaij:2015bpa}, where the $\Delta y$ bins are shifted by $10^{-m}$ for $m=0$, 2, 4, 6 and 8. PROSA 2019 central PDF set is used for predictions with both sets of parameters.
       }
       \label{fig:LHCb}
     \end{figure}
     
Work is in progress \cite{Bai:2021tbd} to assess the PDF and scale uncertainties in the prediction of the forward $(\nu_\tau+\bar{\nu}_\tau)$ flux at the LHC \cite{Bai:2020ukz} using the PROSA 2019 PDF sets as our default PDFs \cite{Zenaiev:2019ktw}.   %
We evaluate the charm meson production cross sections at next-to-leading order (NLO) in perturbative QCD and compare the results with the LHCb data \cite{Aaij:2015bpa} to find optimal input parameters of theoretical calculation. 
Fig. \ref{fig:LHCb} presents the comparison between experimental data and theoretical predictions for $p_T$ distributions of the $D^0+\bar{D}^0$ production cross sections in different rapidity intervals. 
We find that the LHCb data are well-described by the results with renormalization and factorization scales $(\mu_R, \mu_F)$ = (1,~2)$m_T$ and 
the Gaussian smearing parameter $\langle k_T \rangle$ = 1.2 GeV, where the transverse mass $m_T$ equals to $\bigl(p_{T,c}^2 + m_c^2\bigr)^{1/2}$. These are presented with the blue solid histogram. 
Alternative predictions are also shown for input parameters $(\mu_R, \mu_F)$ = (1,1)$m_{T,2}$ and $\langle k_T \rangle$ = 0.7 GeV with $m_{T,2} =  \bigl(p_{T,c}^2 + (2 \ m_c)^2\bigr)^{1/2}$. 
One advantage of the latter input parameter set with respect to the former one is that the optimal value of $\langle k_T \rangle$ is less than 1 GeV, so that it is natural to interpret this parameter as intrinsic transverse momentum for the proton mass scale $\sim$ 1 GeV.
In this work, however, we use the former parameter set that is better matched to the data.

    \begin{figure} [b]
    \centering
       \includegraphics[width=.55\textwidth]{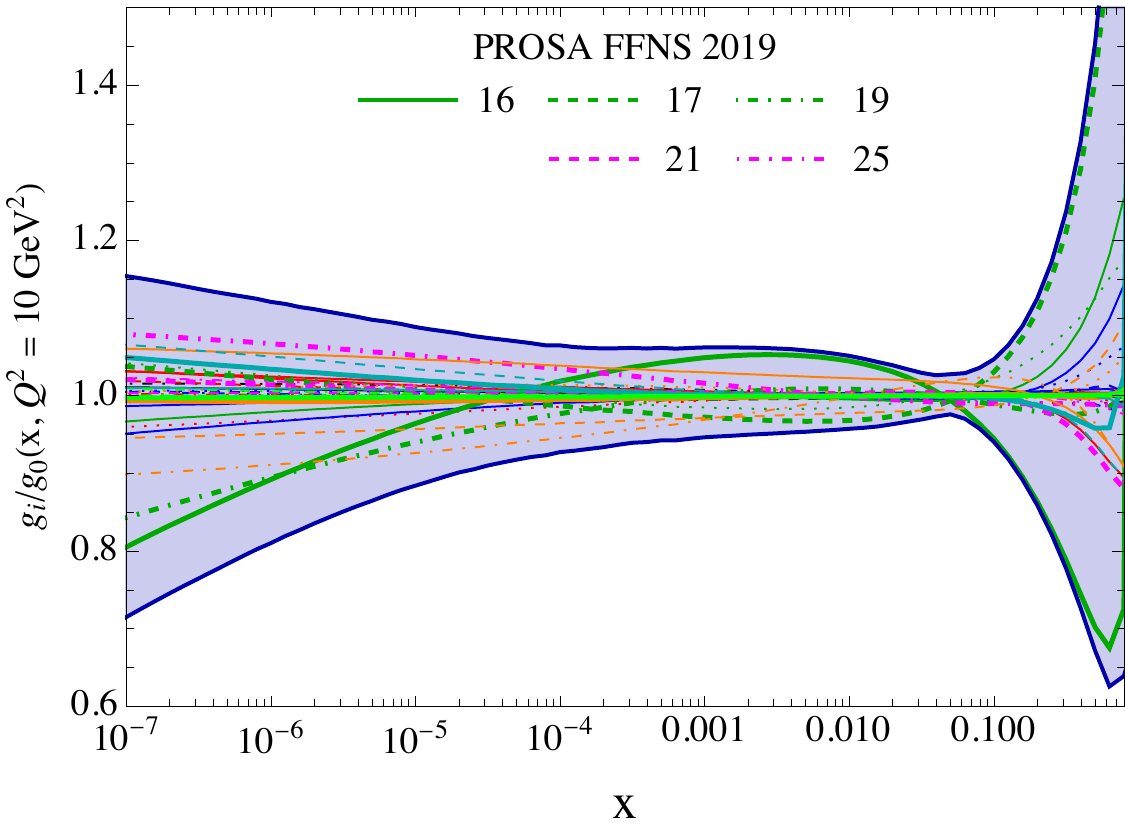}
       \caption{The 40 PROSA FFNS (2019) \cite{Zenaiev:2019ktw}
       sets of gluon distribution function $xg(x,Q^2)$ divided by the best fit set for $Q^2=10$ GeV$^2$. Selected sets are identified by their integer label.}
       \label{fig:pdf-range}
     \end{figure}

Fig. \ref{fig:pdf-range} shows the variation of the 40 sets of gluon distribution function $x\, g (x, Q^2)$, which makes the dominant contribution to the charm production, as function of the parton longitudinal momentum fraction $x$ with $Q^2 = 10 {\ \rm GeV^2 }$.  
Presented are the normalized results to the distribution of the central set, i.e., the best fit.
For the forward production of heavy quark relevant for atmospheric neutrinos, the longitudinal momentum fractions $x$ of the partons involved in the process can be very small for the one from the target and large for the other from the incoming cosmic rays, and reach the $x$ regions where the PDF is currently not well constrained by the experimental data.
Therefore, the uncertainty in both regions is large as shown in the figure.  
In principle, the PDF uncertainty in the prompt flux prediction comes from combining all 40 sets. In this work, we take a simplified approach to approximate the PDF uncertainties.
Among the 40 sets, we select the ones that most deviate from the central set, with particular attention to the small-$x$ and large-$x$ regions, respectively, to approximate the impact of the PDF uncertainty on the atmospheric neutrino flux, discussed in the next section.

\section{Prompt atmospheric neutrino flux}     
     
\begin{figure}[b]
 \centering
       \includegraphics[width=.47\textwidth]{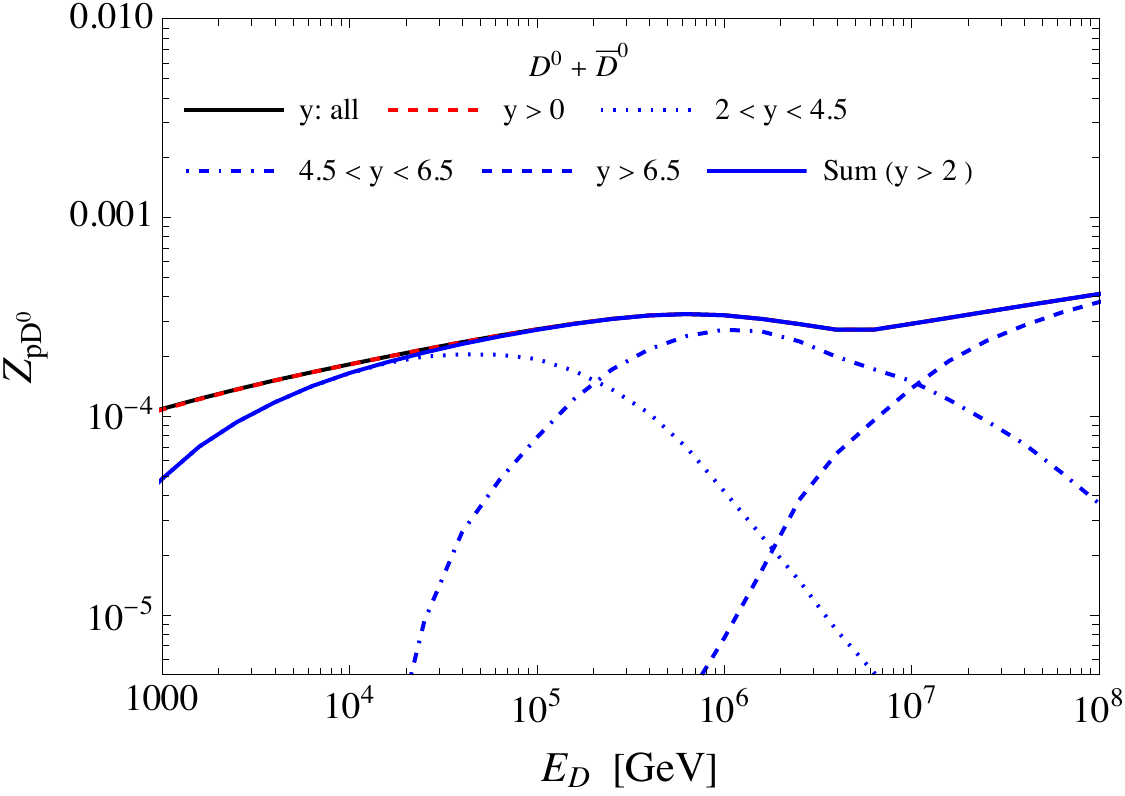}
       \includegraphics[width=.47\textwidth]{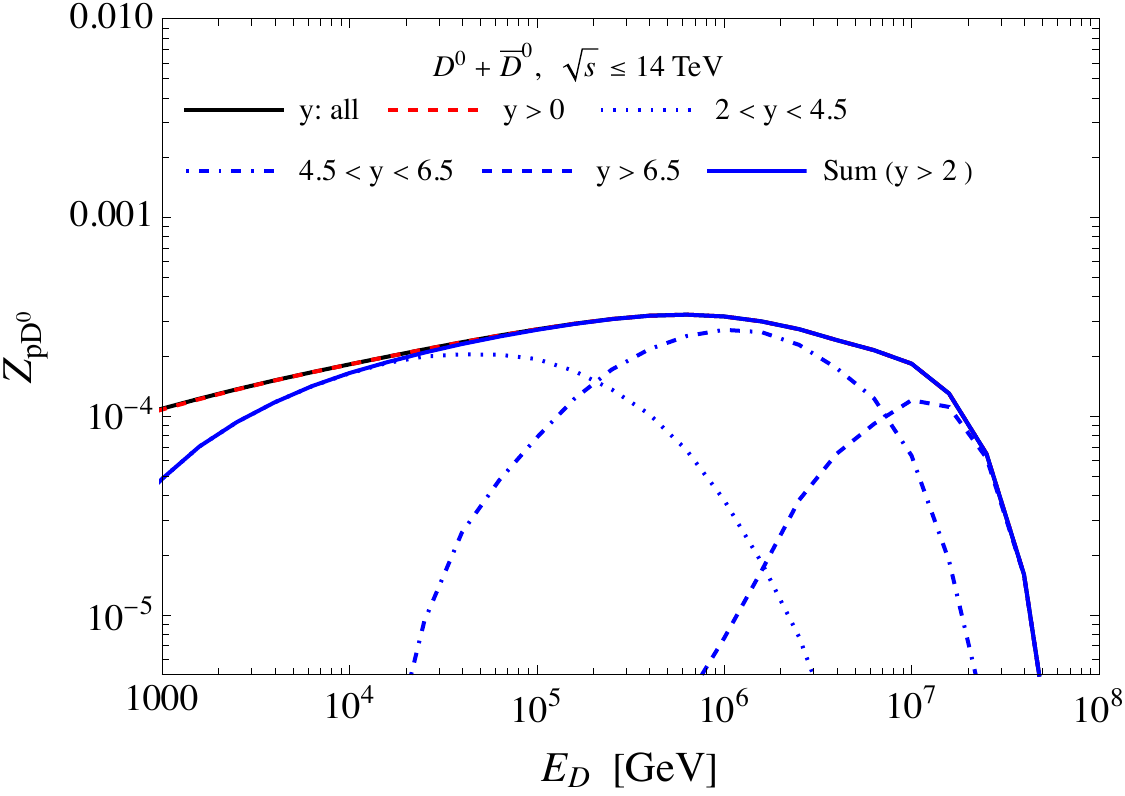}  
       \caption{The production moment $Z_{pD^0}$ for $D^0+\bar{D}^0$
       in proton-Air collisions with a broken power law cosmic ray spectrum.
       Results for all collider frame rapidities $y$ and for $y>0$ overlap in both panels. In the right panel, the $Z$ moment is evaluated with $\sqrt{s}<14$ TeV.}
       \label{fig:zpd0}
\end{figure}

\begin{figure}
 \centering
       \includegraphics[width=.4\textwidth]{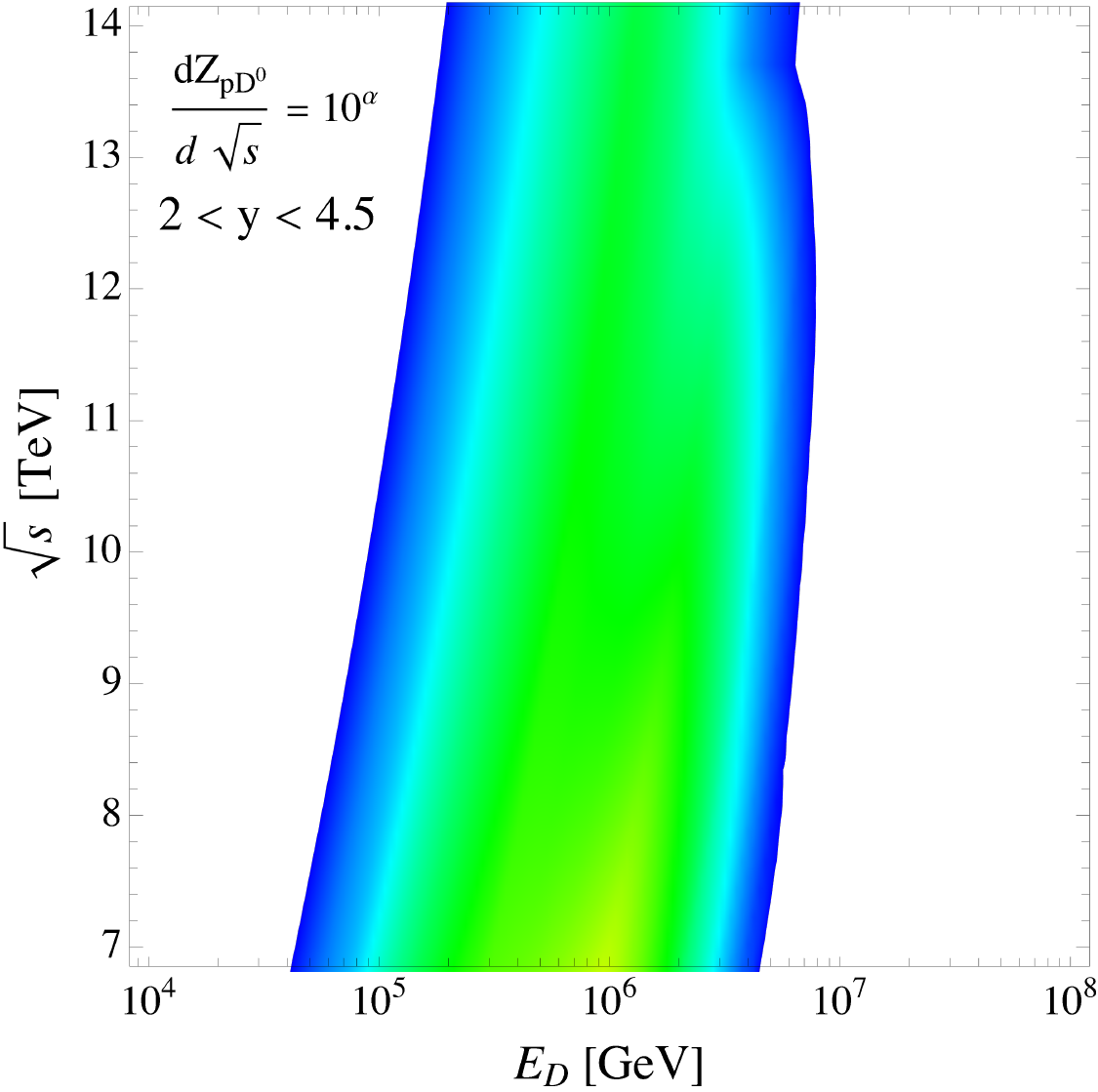} 
       \includegraphics[width=.4\textwidth]{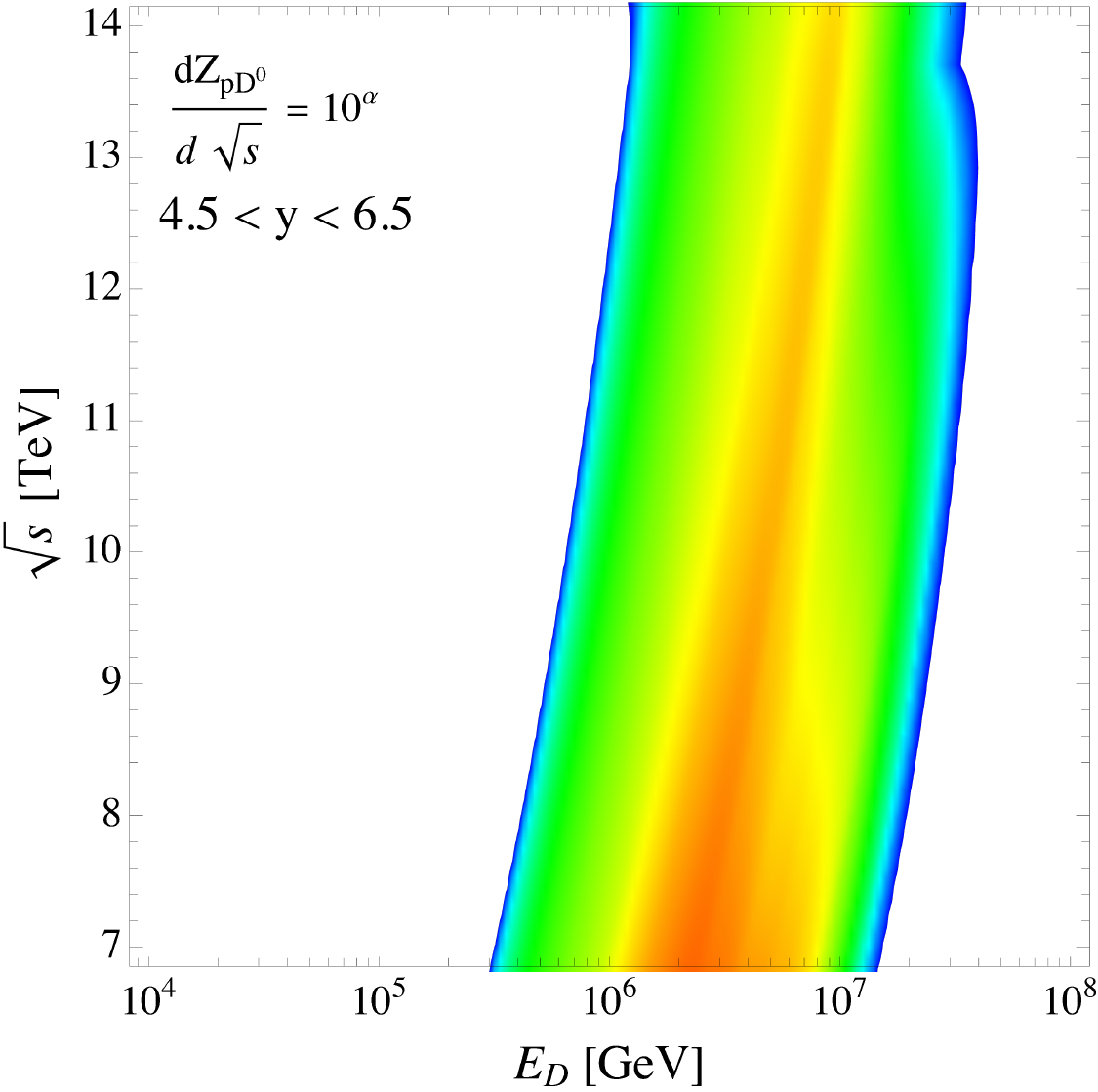}
       \includegraphics[width=.078\textwidth]{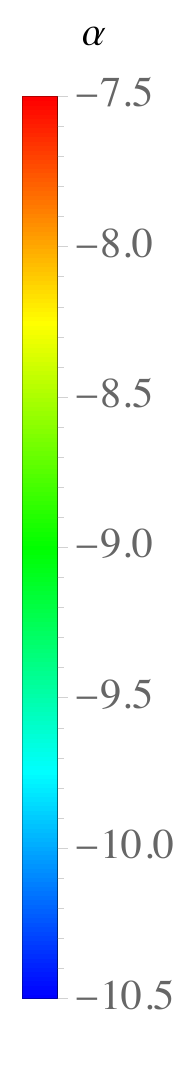}
       \includegraphics[width=.4\textwidth]{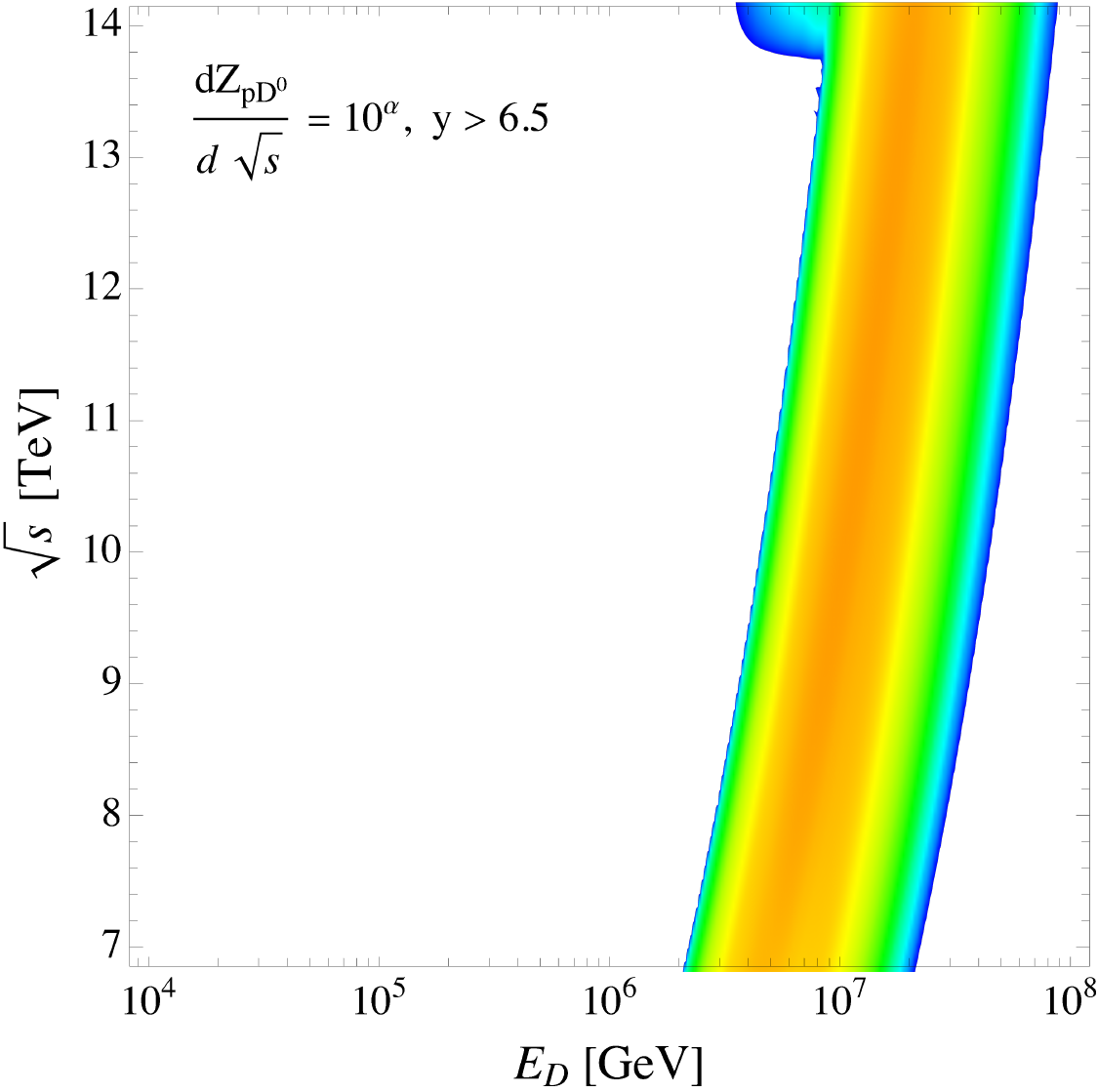} 
       \includegraphics[width=.4\textwidth]{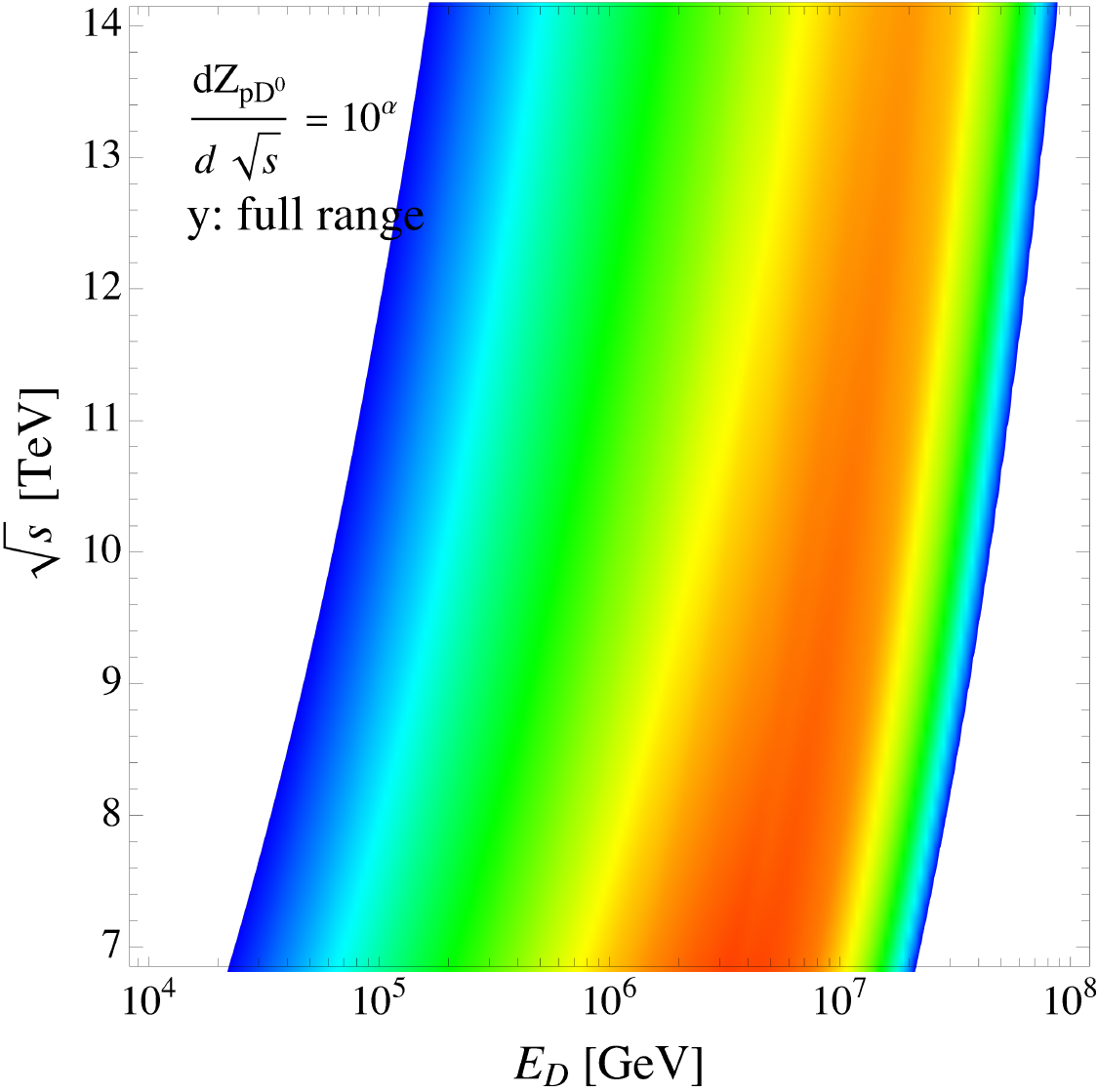}  
       \includegraphics[width=.078\textwidth]{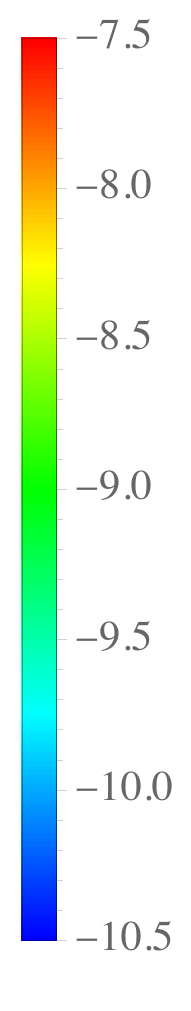}  
       \caption{Distribution of differential $Z_{pD^0}$ in the plane of \{$E_D$, $\sqrt{s}$ \} for different rapidity ranges shown in Fig. \ref{fig:zpd0}. The differential $Z_{pD^0}$ is presented in log scale, i.e. $\alpha = \log_{10} (d Z_{p D^0}/ d  \sqrt{s})$.  } 
       \label{fig:dzpd0-rs}
     \end{figure}
     
The cosmic ray flux is one of the essential factors to evaluate the atmospheric neutrino fluxes. 
Here, we use a broken power law all-nucleon spectrum in terms of nucleon energy $E$ in units of GeV,
that is frequently used to compare predictions from different groups
\cite{Bhattacharya:2016jce,Zenaiev:2019ktw}:
\begin{align}
\label{eq:cr}
\phi_N(E) \, 
[\text{cm}^{-2} \, \text{s}^{-1} \, \text{sr}^{-1} \, (\gev/A)^{-1} ]  
= 
\begin{cases}
1.7 \, E^{-2.7}  \quad  &\text{for } E<5 \cdot 10^6 \,\, \gev \\ 
174 \, E^{-3}           &\text{for } E>5 \cdot 10^6 \,\, \gev\,. 
\end{cases}
\end{align}
The spectrum $\phi_N (E)$ is defined as $\phi_N (E) = dN_N/dE /(dA \, dt \, d\Omega)$ with assumption that the incident cosmic ray flux is isotropic.
Atmospheric neutrino fluxes can be evaluated by the $Z$-moment method, an approximate solution to the coupled cascade equations for protons, hadrons and leptons.
The cascade equation for a particle $j$ is given by 
\begin{eqnarray}
\frac{d\phi_j(E,X)}{dX}&=&-\frac{\phi_j(E,X)}{\lambda_j(E)} - \frac{\phi_j(E,X)}{\lambda^{\rm dec}_j(E,X)}
+ \sum S(k\to j)\,
\end{eqnarray}
with column depth $X$ and interaction (decay) length $\lambda_j^{(\rm dec)}$.
The source term $S (k\to j)$ can be approximated by rewriting it in terms of a $Z$ moment, 
\begin{eqnarray}
S(k\to j) &\simeq & Z_{kj}(E)\frac{\phi_k(E,X)}{\lambda_k(E)}\, ,
\end{eqnarray}
which, in turn, can be calculated using as input the cosmic ray flux, the interaction (or decay) length and the differential production cross section (or differential decay distribution), as a function of the energy, 
\begin{eqnarray}
Z_{kj}(E) &=& \int _E^{\infty}dE ' \frac{\phi_k^0(E')}{\phi_k^0(E)}
\frac{\lambda_k(E)}{\lambda_k(E')}
\frac{dn(k\to j;E',E)}{dE} \, ,
\end{eqnarray}
for $\phi_k (E,X)\simeq \phi_k^0(E)\exp{(-X/\Lambda_k)}$.
Here, the energy distribution of the particles produced by interaction or decay, $dn(k\to j)/dE$ is given by
\begin{eqnarray}
\frac{dn(k\to j;E',E)}{dE}&=&
\begin{cases}
\frac{1}{\sigma_{kA}(E')}\frac{d\sigma(kA\to jY;E',E)}{dE }
\quad {\rm (interaction)\, ,}\\
\frac{1}{\Gamma_k(E')}\frac{d\Gamma (k\to jY;E',E)}{dE}\quad {\rm (decay)}
\ .
\end{cases}
\end{eqnarray}
In order to evaluate the prompt neutrino flux, 
the $Z$ moments for proton regeneration, $Z_{pp}$,
and hadron production, $Z_{ph}$ and $Z_{hh}$, are required as well as $Z_{h\nu}$ for decays to neutrinos, where $h$ denotes hadron. 
The calculation of $Z$ decay moments is relatively straightforward.
On the other hand, the $Z$ production moments, especially $Z_{ph}$ for hadron production bring about large uncertainty in the prediction of prompt atmospheric neutrino flux. The $Z$ moments can be 
partly constrained  by new measurements at the Forward Physics Facility at the LHC.

In Fig. \ref{fig:zpd0}, we show the contributions to the $Z$ production moment $Z_{pD^0}$ for different center of mass rapidity regions. The $Z$ moment includes a factor of $2$ for the production of both $D^0$ and $\bar{D}^0$ mesons.
The predictions in the right panel are evaluated with $\sqrt{s}\leq14$ TeV in the cosmic ray-air interactions modeled with proton-nucleon collisions. (See also ref. 
\cite{Goncalves:2017lvq}.)
The plot shows that charm hadron production with the highest LHC energy $\sqrt{s}=14$ TeV impacts the prompt neutrino flux prediction for neutrino energies up to a few 10's of PeV, which implies that LHC measurements can contribute to our understanding of the prompt atmospheric neutrino flux below $E_\nu \sim 10^7$ GeV.
Our result also indicates that prompt atmospheric neutrinos at such energies are from the charm hadrons produced mainly in the rapidity of $y \lesssim 6.5$ but also include higher rapidities.

Fig. \ref{fig:dzpd0-rs} presents the distribution of the differential Z moments for $D^0+\bar{D}^0$ production in the plane of \{$E_D, \sqrt{s}$ \} for the different rapidity ranges. 
The prompt atmospheric neutrinos become important in the energy range of $E \gtrsim 10^5 - 10^6$ GeV, where they start dominating the conventional atmospheric neutrino flux. 
In the plots of Fig. \ref{fig:dzpd0-rs}, one can see the range of both collision energy and rapidity that make important contribution to production of charm hadrons for prompt neutrinos in the energy range of interest.

 \begin{figure}
    \centering
       \includegraphics[width=.47\textwidth]{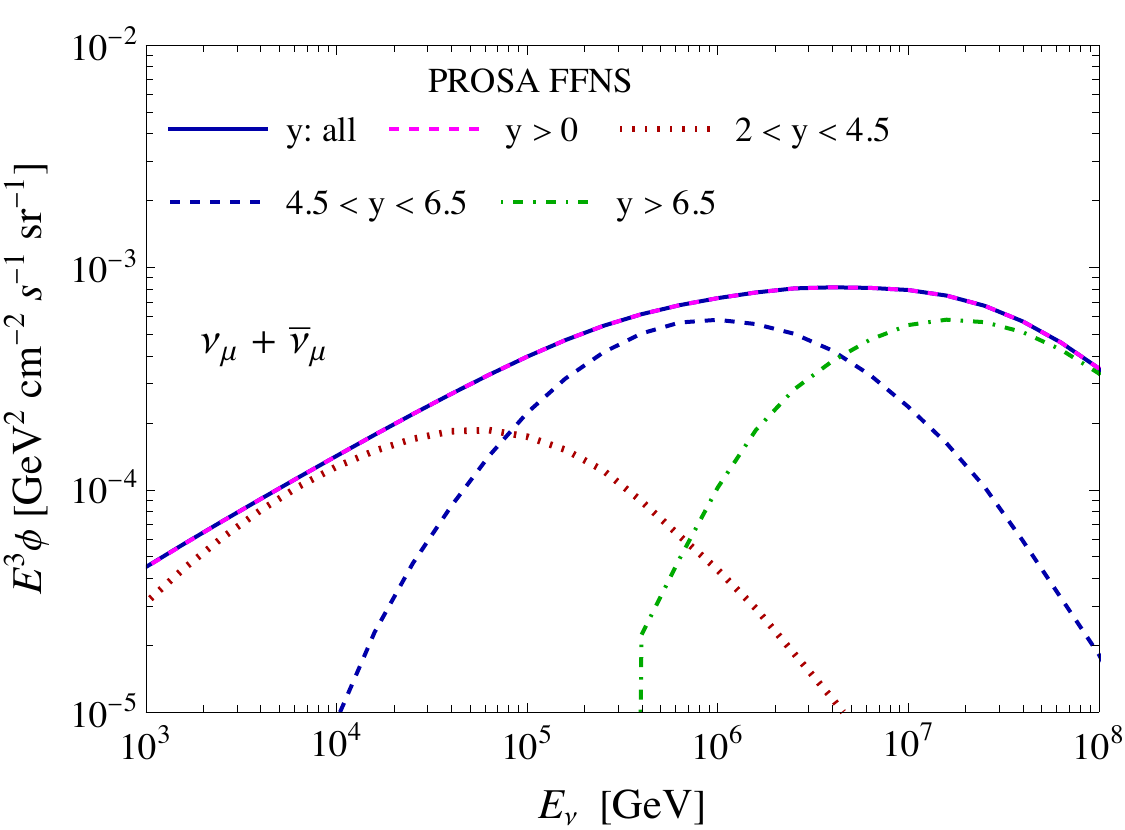}
       \includegraphics[width=.47\textwidth]{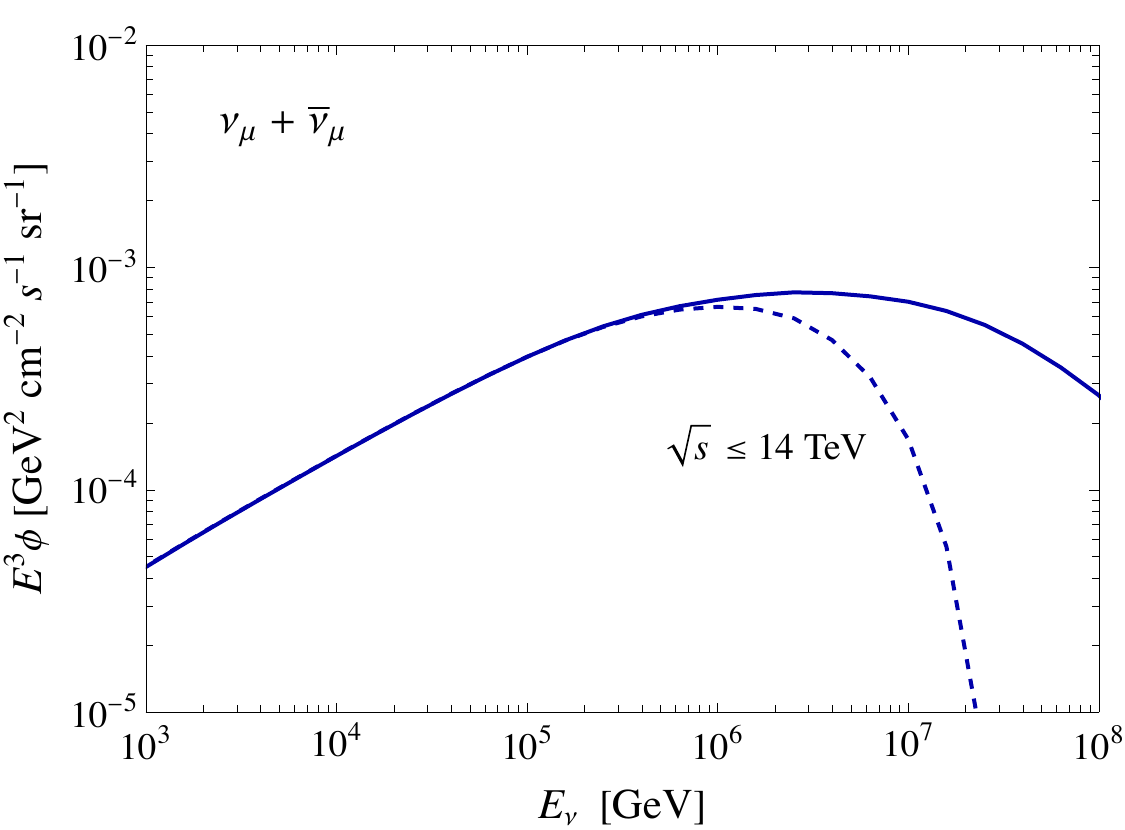}      
       \caption{The prompt flux of atmospheric $\nu_\mu+\bar{\nu}_\mu$ from different charm meson rapidity ranges in
       $pp\to c\bar{c}X$, using a broken power law cosmic ray spectrum (left). 
       The right panel is the flux evaluated with $\sqrt{s}<14 {\ \rm TeV}$.}
       \label{fig:e3phi}
     \end{figure}

\begin{figure} 
     \centering
       \includegraphics[width=.49\textwidth]{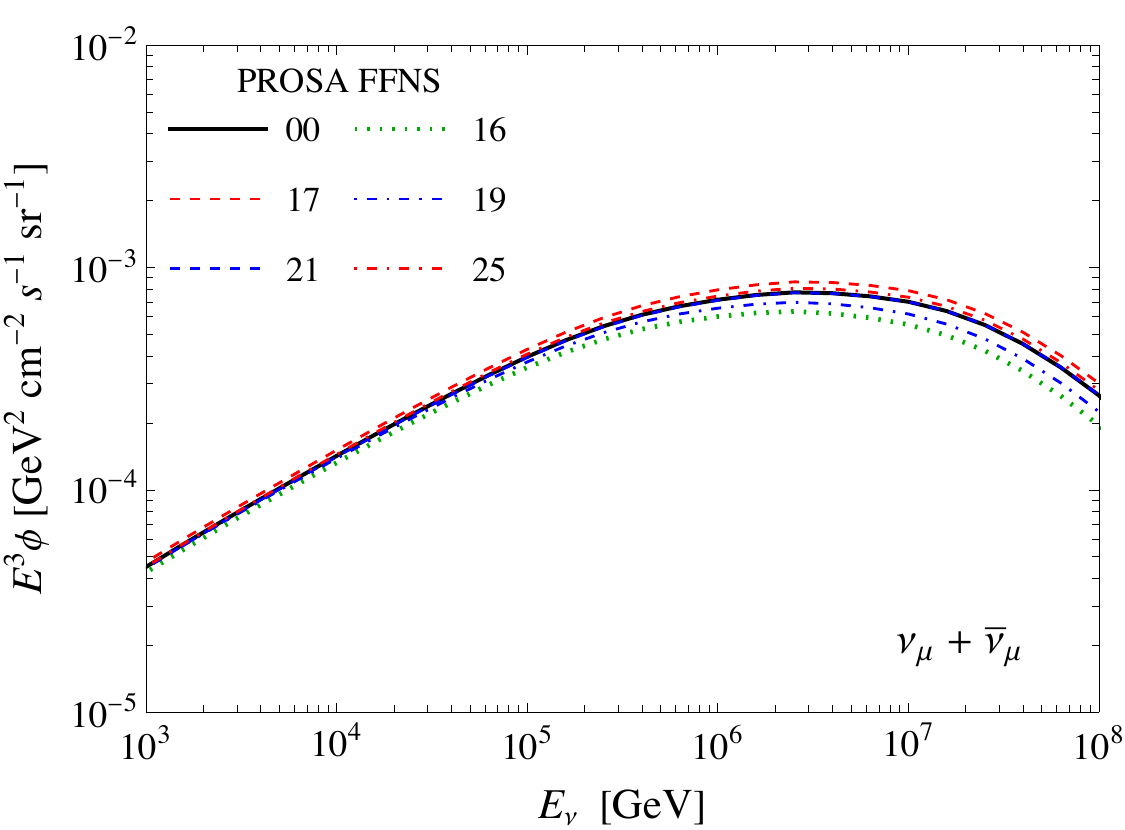}
       \includegraphics[width=.47\textwidth]{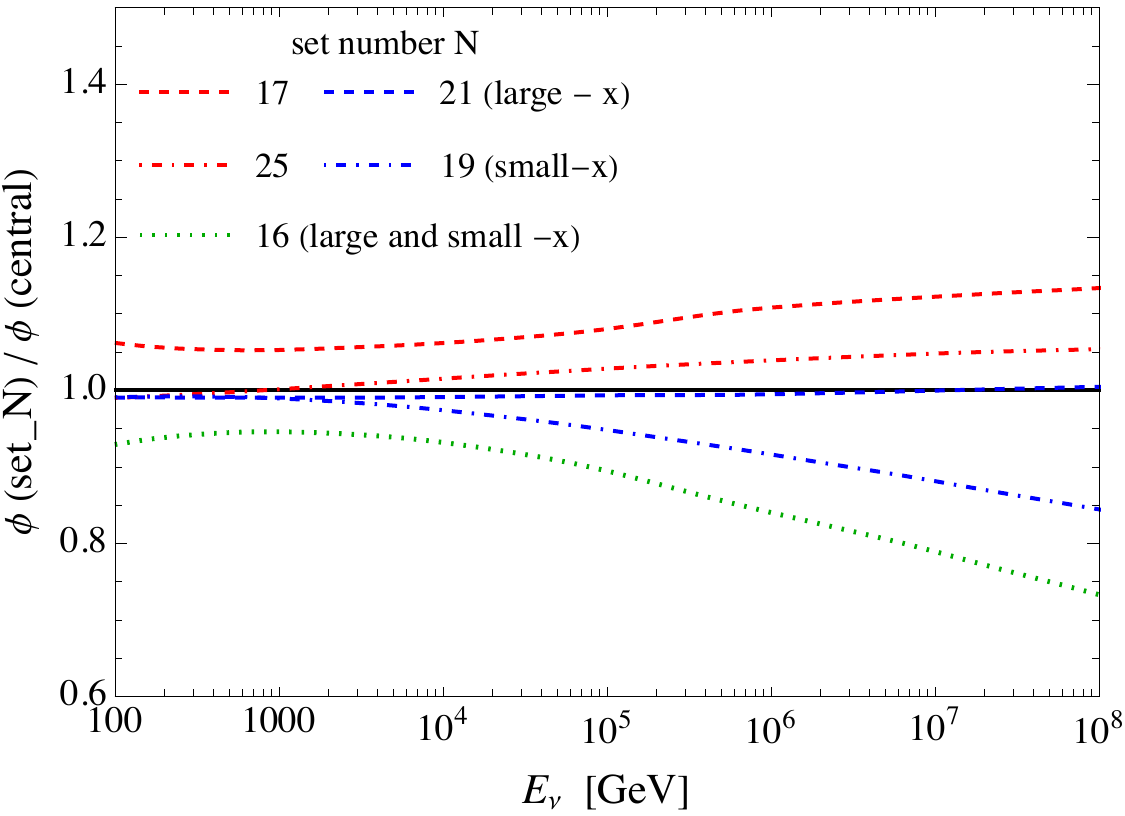}       
       \caption{The prompt fluxes of atmospheric $\nu_\mu+\bar{\nu}_\mu$ evaluated with the different PDF sets of PROSA FFNS (2019) selected from Fig. \ref{fig:pdf-range} (left) and their ratio to the central set (right).  }
       \label{fig:e3phi-x}
     \end{figure}

The resulting flux of prompt atmospheric neutrinos is shown in fig. \ref{fig:e3phi}, evaluated with the production and decay moments that involve the $D^0+\bar{D}^0$, $D^\pm$, $D_s^\pm$ and $\Lambda_c^\pm$ hadrons. 
The charm hadron flux in the high energy and low energy limits are used to evaluate the respective decay moments. Together, these moments and the assumed cosmic ray flux from eq. (\ref{eq:cr}) are used to evaluate the prompt neutrino flux. Details are provided in refs. \cite{Bhattacharya:2016jce,Zenaiev:2019ktw} and references therein. For definiteness, we show the prompt $\nu_\mu+\bar{\nu}_\mu$ atmospheric flux, which is equal to the prompt $\nu_e+\bar{\nu}_e$ atmospheric flux.

IceCube events in the 10's of TeV are dominated by the conventional neutrino flux. At the higher energy range of IceCube events, the prompt neutrino flux lies below the astrophysical neutrino flux \cite{IceCube:2020wum,Schukraft:2013ya}. Prompt atmospheric neutrinos in the few PeV range come mainly from the charm hadron rapidity region in the collider frame of $4.5-6.5$.

Fig. \ref{fig:e3phi-x} presents the variation of the prompt atmospheric neutrino flux predictions according to the PDF set members selected in fig. \ref{fig:pdf-range} for maximally deviated ones from the best fit at the respective small $x$ and large $x$ region (left panel). Their  ratio to predictions with the central set is shown in the right panel of fig.
\ref{fig:e3phi-x}. The impact of the PDF uncertainty is within 30 \% and is largest at the highest energy.

\section{Discussion}
We have investigated the possibility of interplay between the LHC and the atmospheric interaction for prompt neutrino production.
We have explored the effect of center-of-mass collision energy and the rapidity that could be accessible at the LHC on charm hadron production in the atmosphere and impact on the fluxes of prompt atmospheric neutrinos. 
The prompt atmospheric neutrinos that are important as backgrounds to astrophysical neutrinos in the few PeV energies are mainly from the charm hadrons produced in the rapidity of $4.5 < y < 6.5$ with additional contributions from $y>6.5$. 
Up to now, the LHCb experiment measured charm production at $2 < y < 4.5$. The new forward experiments, FASER$\nu$ and SND@LHC will cover $y \gtrsim 8.8$ and $7.2 < y < 8.6$, respectively. 
Therefore, if the charm hadron production can be probed in the rapidity range $4.5 < y < 7.2$  at future Forward Physics Facility, it will be able to provide relevant new constraints for the theoretical predictions of the prompt atmospheric neutrino flux.

In forward charm production in $pp$ collisions, small and large values of the parton momentum fraction $x$ are involved.
Measurements at LHCb constrain PDFs as low as $x\sim 10^{-6}$ \cite{Zenaiev:2019ktw,Gauld:2016kpd}. Experiments in the more forward region at the LHC will measure forward production of prompt neutrinos, sensitive to even smaller values of $x$.
The large value of $x$ is approximately related to rapidity according to $x\sim (m_T/\sqrt{s}) e^{ y}$. At the LHC with
$\sqrt{s}=14 \ {\rm TeV}$, $x\sim 0.01\ (0.1)$ for $y=4.5 \ (6.5)$. PDF uncertainties are only part of the larger uncertainties associated with perturbative QCD evaluations of the prompt atmospheric neutrino fluxes \cite{Bhattacharya:2016jce,Zenaiev:2019ktw}. Measurement of charm production and their associated neutrinos in new kinematic regimes will guide future theoretical predictions of the prompt atmospheric neutrino flux.

\acknowledgments
This work is supported in part by U.S. Department of Energy Grants DE-SC-0010113 and DE-SC-0012704, the National Research Foundation of Korea (NRF) grant funded by the Korea government(MSIT \footnote{MSIT : Ministry of Science and ICT}) (No. 2021R1A2C1009296) and by the german Bundesministerium f\"ur Bildung und Forschung (contract 05H18GUCC1).

%
%
%

\end{document}